\documentclass[10pt,twocolumn,twoside]{ieeeconf}
  
\IEEEoverridecommandlockouts    
\usepackage{amsthm}
\newtheorem{remark}{Remark}

\newtheorem{lemma}{Lemma}
\newtheorem{proposition}{Proposition}

\newtheorem{theorem}{Theorem}
\newtheorem{assumption}{Assumption}

\newtheorem{scenario}{Scenario}

\usepackage{algorithm}
\usepackage[noend]{algpseudocode}
\usepackage{cite}
\usepackage{amsmath}
\usepackage{caption}
\usepackage{amstext}
\usepackage{amssymb}
\usepackage{tikz}
\usepackage[mathscr]{euscript}
\usepackage{color}
\usepackage{amsfonts}
\usepackage{arydshln}

\usepackage{svg}
\usepackage{siunitx}

\usepackage{cite}

\let\labelindent\relax

\usepackage[final]{pdfpages}
\usepackage{csquotes}
\makeatletter
\newcommand*{\rom}[1]{\expandafter\@slowromancap\romannumeral #1@}
\makeatother
\usepackage{graphicx}      

\usepackage{xparse}
\usepackage{mathdots}
\usepackage{epstopdf}

\newcommand{\seb}[1]{%
{\leavevmode\color{black}#1}%
}

\newcommand{\sebcancel}[1]{%
{\leavevmode\color{black}#1}%
}

\newcommand{\lud}[1]{%
{\leavevmode\color{red}#1}%
}

\newcommand{\cha}[1]{%
{\leavevmode\color{olive}#1}%
}

\DeclareMathAlphabet\mathbfcal{OMS}{cmsy}{b}{n}
\usepackage{array}

\usepackage{comment}

\usepackage{mathtools,xparse}

\DeclarePairedDelimiter{\abs}{\lvert}{\rvert}
\DeclarePairedDelimiter{\norm}{\lVert}{\rVert}

\usepackage{hyperref}
\usepackage{pdfpages}
\usepackage{xcolor}
\usepackage{subcaption}
\usepackage{enumitem}

\begin{document}
\author{Ludvig Lindström, Sebin Gracy, Sindri Magnússon, and Henrik Sandberg
\thanks{Ludvig Lindström, Sebin Gracy  and Henrik Sandberg are with the Division of Decision and Control Systems, School of Electrical Engineering  and  Computer  Science,  KTH Royal Institute of Technology, and Digital Futures, Stockholm, Sweden. Emails:lulinds@kth.se, gracy@kth.se, hsan@kth.se. This work has received funding from the DEMOCRITUS project.\\
Sindri Magnússon is with the Department of Computer and Systems Sciences at Stockholm University. Email:sindri.magnusson@dsv.su.se
}}
\title{Leakage Localization in Water Distribution Networks: \\ A Model-Based Approach}
\maketitle
\begin{abstract}
 The paper studies the problem of 
 leakage localization in water distribution networks. 
 For the case of a single pipe that suffers from a single leak, by taking recourse to pressure and flow measurements, and assuming those are noiseless, we provide a closed-form expression for leak localization, leak exponent and leak constant. For the aforementioned setting, but with noisy pressure and flow measurements, an expression for estimating the location of the leak is provided. Finally, assuming the existence of a single leak, for a network comprising of more than one pipe and assuming that the network has a tree structure, we provide a systematic procedure for determining the leak location, the leak exponent, and the leak constant.
\end{abstract}

\section{Introduction}


The United Nations recognises that  access to clean water is a fundamental human right. 
Nevertheless, water is scarce in many parts of the world, and according to the United Nations, as many as $785$ million people lack basic drinking-water services today. Moreover, due to climate change, it is estimated that half of the world's population will soon be living in water-stressed areas \cite{UNwater}. At the same time, massive amounts of water are wasted because of leakages in (aging) water distribution infrastructures.\footnote{The term \enquote{leakages} is also referred to as \enquote{unaccounted-for-water} elsewhere in the literature; see for instance \cite{lai1991unaccounted}.} 
%
For example, according to recent estimates, in England and Wales, as much as $3$ billion liters of water is lost every day due to leakage, which is roughly one third of the daily domestic use \cite{UKwater}. In many other countries,  it is estimated that 20\% to 30\% of all water is lost due to leakage~\cite{adedeji2017towards,chan2018review}. Leakages due to broken pipes also threaten the health of the population accessing water through such pipes   as they give easy access to bacteria and soil to possibly contaminate the drinking water. Moreover,  soil and water samples tested in the neighborhood of sources of water 
at eight locations in six US states found that these soils often contain potentially harmful bacteria and pathogens \cite{karim2003potential}. Yet another consequence of leakage is the corresponding reduction in the revenue that the concerned utility incurs \cite{lai1991unaccounted}.

 Leakages are difficult to localize, since water distribution networks are mostly underground and, therefore, difficult to reach. Fixing  leakages is non-trivial from an economic standpoint since it requires digging into the ground, rupturing roads and potentially inconveniencing other urban infrastructures. Moreover, if the location of the leakages is not known, then it is practically impossible to fix them. 
 Therefore, good leakage localization methods are essential for the sustainability of our water distribution networks. Most of the existing practices for localizing leakages are slow and costly, as these require technicians to go around the system and test different places for leakages and broken pipes. \sebcancel{In recent times, an approach based on the frequency response for fault identification in pipelines has been provided in \cite{rubio2017blockage}. The approach therein, besides leak identification, also accounts for simultaneous detection of blockages in pipelines. Certain other approaches have been presented in, among others, \cite{wang2018pipeline,wang2019spectral}.} 
 
  Digitalization of 
 water distribution infrastructures can bring a huge promise for improving leakage localization. There are already ongoing initiatives to introduce smart water meters at the end-users to help monitor water usage~\cite{booth2018digital}. It is expected that in the future there will be diverse sensors throughout the water distribution infrastructures to improve their safety and efficiency, e.g., by detecting and localizing leakages and broken pipes. 
 In fact, leakage localization from sensor readings has already started to attract attention in the literature, see, e.g., the recent reviews~\cite{adedeji2017towards,chan2018review}  and references therein. Most existing approaches use active localization techniques \cite{lee2007leak,leandro2009comparison,kafle2020active}. That is, to localize a leakage, some signals (e.g., water flow at particular frequencies or sound waves) are injected into the pipes with the sole purpose of facilitating localization. Moreover, these approaches often need some human involvement, e.g., to find a place to inject the signal into the system. 
 
 For practical impact, it would be more attractive to use passive localization approaches, i.e., those that do not 
 interfere with the system operations. There have been some nice efforts  in this direction~ \cite{kang2017novel,hu2020tngan}. An optimization-based approach that relies on pressure and flow measurements to detect leaks has been proposed recently in \cite{vrachimis2021leakage}. 
  \sebcancel{Observe that the approach therein is fully data-driven}, and thus they do not exploit the rich structures attained from physical models to facilitate localization. In fact, water distribution systems are governed by well-grounded laws of physics, that can provide a lot of information about water flow and potential leaks in these systems. In the present paper, we focus on such an approach for leak localization.




The main contribution of this paper is to illustrate how physical models of flow in water distribution networks can be used to localize leakages. 
More concretely, our contributions can be summarized as follows:
\begin{enumerate}[label=\roman*)]
    \item \label{q1} Considering a water distribution system with a single pipe, and suffering from a single leak, under the assumption of perfect pressure and flow measurements, we provide a closed-form expression for locating the position of the leak, the leak exponent, and the leak constant; see Theorem~\ref{th:sen1compile}.
    \item \label{q2} For the same setting as in \ref{q1}, but assuming that the pressure and flow measurements are possibly corrupted by noise, we provide an expression for estimating the location of the leak; see Theorem~\ref{th:sen1Noise}.
    \item We consider a water distribution system comprising of more than one pipe, but with the topology of the underlying network being a tree graph, and suppose that the system suffers from a single leak. Under the assumption of perfect pressure and flow measurements, we provide a systematic procedure for determining the location of the leak, the leak exponent, and the leak constant; see Theorem~\ref{th:TreeIsolate}.
\end{enumerate} \vspace{-4mm}

\subsection*{Paper Outline}
We conclude this section by collating all the notations needed in the sequel. The model for the water distribution network and the background material needed are provided in Section~\ref{sec:prelims}. Our main results are spread over Section~\ref{sec:single pipe} and~\ref{sec:tree}. In particular, in Section~\ref{sec:single pipe} we study a water distribution network consisting of a single pipe and suffering from a single leak, whereas in Section~\ref{sec:tree} we consider water distribution networks comprising of more than one pipe, the underlying network structure being a tree graph, and suffering from a single leak. Our theoretical findings are illustrated in Section~\ref{sec:simulations}. Finally, we summarize the paper, and highlight some research directions of possible interest to the wider research community in Section~\ref{sec:conclusions}. All proofs are available at \cite{ludvigECC}.

\subsection{Notations and Graph-theory terminology}\label{sec:notations}
Let $\mathbb{Z}$ denote the set of all integers. An ordered set, $(n,\dots,m)$, is called a sequence. For $n,m\in \mathbb{Z}$,  $(k)_{k=n}^{m}=(n,{n+1},\dots, {m-1},{m})$ denotes a sequence.

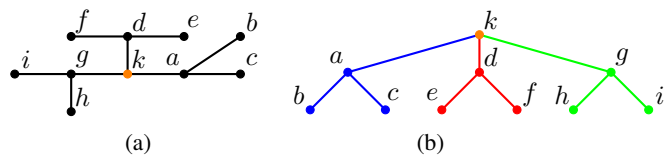
\begin{figure}
 \begin{subfigure}{.45\linewidth}
     \centering

\begin{tikzpicture}[scale=0.5]

    \coordinate (HT) at (0,-1);
    
    \node at (0.3, -0.6) {$k$};
    \node at (-1.2, -0.6) {$g$};
    
    \coordinate (HL) at (1.5,-1);
    \coordinate (HM) at (-1.5,-1);
    \coordinate (HR) at (0,0);
    
    \coordinate (HLL) at (3,0);
    \coordinate (HLR) at (3,-1);


    \coordinate (HML) at (-3,-1);
    \coordinate (HMR) at (-1.5,-2);
    
    
    \coordinate (HRL) at (1.5,0);
    \coordinate (HRR) at (-1.5,0);
    
    \node at (1.2, -0.6) {$a$};
    \node at (0.3, 0.4) {$d$};
    
    \node at (3.3, 0.4) {$b$};
    
    \node at (3.3, -0.6) {$c$};
    
    
    \node at (-1.2, -1.6) {$h$};
    ;
    \node at (1.75, 0.4) {$e$};
    
    \node at (-1.2, 0.4) {$f$};
    
    
     \node at (-2.7, -0.6) {$i$};
    
    
    \draw[thick] (HL) to (HT);
    \draw[thick] (HM) to (HT);
    \draw[thick] (HR) to (HT);
    
     \draw[thick] (HL) to (HLL);
     \draw[thick] (HL) to (HLR);
    
    
    \draw[thick] (HM) to (HML);
    \draw[thick] (HM) to (HMR);
    
    
    \draw[thick] (HR) to (HRL);
    \draw[thick] (HR) to (HRR);

    \filldraw[orange] (HT) circle (3pt);
    
    \filldraw[black] (HL) circle (3pt);
    \filldraw[black] (HM) circle (3pt);
    \filldraw[black] (HR) circle (3pt);
    
    \filldraw[black] (HLL) circle (3pt);
    \filldraw[black] (HLR) circle (3pt);
    
    
    \filldraw[black] (HML) circle (3pt);
    \filldraw[black] (HMR) circle (3pt);
    
    
    \filldraw[black] (HRL) circle (3pt);
    \filldraw[black] (HRR) circle (3pt);
    
    
    
    
    
    

    
    
    
    
    
    
    
\end{tikzpicture}
      \subcaption{}
      \label{fig:left}
     \end{subfigure}%
     \begin{subfigure}{.45\linewidth}
    \centering

\begin{tikzpicture}[scale=0.5]

    \coordinate (T) at (0,0);
    
    \coordinate (L) at (-3.5,-1);
    \coordinate (M) at (0,-1);
    \coordinate (R) at (3.5,-1);
    
    \coordinate (LL) at (-4.5,-2);
    \coordinate (LR) at (-2.5,-2);

    
    \coordinate (ML) at (-1,-2);
    \coordinate (MR) at (1,-2);
    
    
    \coordinate (RL) at (2.5,-2);
    \coordinate (RR) at (4.5,-2);
    
    \node at (0.3, 0.4) {$k$};
    
    \node at (-3.8, -0.6) {$a$};
    \node at (0.3, -0.6) {$d$};
    \node at (3.8, -0.6) {$g$};
    
    \node at (-4.8, -1.6) {$b$};
    \node at (-2.3, -1.6) {$c$};
    
    \node at (-1.25, -1.6) {$e$};
    \node at (1.35, -1.6) {$f$};
    
    \node at (2.2, -1.6) {$h$};
    \node at (4.8, -1.6) {$i$};

    
    \draw[thick,blue] (L) to (T);
    \draw[thick,red] (M) to (T);
    \draw[thick,green] (R) to (T);
    
    \draw[thick,blue] (L) to (LL);
    \draw[thick,blue] (L) to (LR);
    
    
    \draw[thick,red] (M) to (ML);
    \draw[thick,red] (M) to (MR);
    
    
    \draw[thick,green] (R) to (RL);
    \draw[thick,green] (R) to (RR);

    \filldraw[orange] (T) circle (3pt);
    
    \filldraw[blue] (L) circle (3pt);
    \filldraw[red] (M) circle (3pt);
    \filldraw[green] (R) circle (3pt);
    
    \filldraw[blue] (LL) circle (3pt);
    \filldraw[blue] (LR) circle (3pt);
    
    
    \filldraw[red] (ML) circle (3pt);
    \filldraw[red] (MR) circle (3pt);
    
    
    \filldraw[green] (RL) circle (3pt);
    \filldraw[green] (RR) circle (3pt);
    
    
    




    
    

\end{tikzpicture}   
    \subcaption{}
    \label{fig:right}
      
     \end{subfigure}
     \caption{\seb{(a) A tree. (b) The same tree as in (a), where the blue, red, and green trees are the subtrees of vertex $k$ and where $k$ is present in all three subtrees.}}
     \label{fig:mylabel}
 \end{figure}

\sebcancel{Consider a (possibly) directed graph $\mathcal G$ with $n$ vertices. Let $A$ be the matrix that represents the interconnections between the various vertices. That is, $A_{ij} \neq 0$ if there is an edge from vertex $j$ to vertex $i$; otherwise, $A_{ij} =0$.
Let $\mathcal V=\{1,2,\hdots, n\}$ denote the set of vertices of $\mathcal G$. Let $\mathcal E=\{(i,j)\mid A_{ij}\neq 0\}$ denote the set of edges of $\mathcal{G}$. We say that vertex $j$ is a neighbor of vertex $i$ if $A_{ij} \neq 0$.}
For $\mathcal{G}=(\mathcal{V},\mathcal{E})$, let $\mathcal{V}_i=\{a\in\mathcal{V}|(a,i)\in \mathcal{E}\}$ denote the set of vertices that are neighbors of vertex $i$. For a vertex $a\in\mathcal{V}$, $\abs{\mathcal{V}_a}$ is called the degree of $a$, and is denoted as $\delta(a)$. A graph $\mathcal{G}'=(\mathcal{V}',\mathcal{E}')$ is a subgraph of a graph $\mathcal{G}=(\mathcal{V},\mathcal{E})$ if $\mathcal{V}'\subseteq \mathcal{V}$ and $\mathcal{E}'\subseteq \mathcal{E}$. For a graph $\mathcal{G}$, a sequence of vertices $(k)_{k=n}^{m}$ for which $\forall k$, vertices $k$ and ${k+1}$ are \sebcancel{neighbors}, 
is called a walk. A walk where all vertices are distinct is called a path. A walk $(k)_{k=n}^{m}$ where $(k)_{k=n}^{m-1}$ is a path and $n=m$ is called a cycle. If for every two vertices $a,b\in \mathcal V$ there exists a path $(k)_{k=a}^{b}$, then $\mathcal{G}$ is called a connected graph. A tree, $\mathcal{T}=(\mathcal{V},\mathcal{E})$, is a graph that is connected and in which there are no cycles. A vertex $a\in \mathcal{V}$, for which $\delta(a)=1$, is called a leaf. For $\mathcal{T}=(\mathcal{V},\mathcal{E})$, let $\mathcal{L}=\{a\in \mathcal{V}|\delta(a)=1 \}$ be the leaf set. A path between two vertices $a,b\in \mathcal{V}$ in a tree is unique \cite{biggs,subgraph}. 
\seb{Consider a tree $\mathcal{T}$, pick a vertex $m \in \mathcal{V}$, and call this vertex as the \emph{root} vertex. For any vertices $p, q \in \mathcal V$, we say that $q$ is a descendant of $p$ if $p\in(k)_{k=m}^q$. We say that a subgraph $\mathcal{T'}(\subseteq \mathcal{T})$ is a subtree if $\mathcal{T}'$ is a tree. 
Given a vertex $m \in \mathcal V$, in the sequel, we will be interested in considering \emph{all} subtrees $\mathcal T'$ that are rooted at $m$\footnote{We say a subtree is rooted at a vertex $m$ if every vertex $p \in \mathcal V$ is a descendant of $m$.}, and furthermore, $m\in\mathcal{V}'$, for every $a,b\in\mathcal{V}'\setminus\{m\}$, $m\notin(k)_{k=a}^b$, and all descendants of $a$ ($a \neq m$) are included in $\mathcal{V}'$; for a pictorial depiction see Figure~\ref{fig:mylabel}.}



\section{Preliminaries}\label{sec:prelims}

\subsection{Water Distribution Systems}\label{sec:WaterDistributionSystems}

We consider a water distribution system with $F$ vertices represented by the \seb{graph} $\mathcal{G}=(\mathcal{V},\mathcal{E})$. Throughout this paper, we assume that the graph is a tree; we classify the vertices as leaf vertices and non-leaf vertices. 
We denote by $\mathcal{L}\subseteq \mathcal{V}$ the leaf  vertices. 
The vertices $i\in \mathcal{V} \setminus \mathcal{L}$ represent junctions. The edges $(i,j)\in \mathcal{E}$ represent a pipe between the vertices $i$ and $j$.

The fluid dynamics model governing a closed pipe water distribution system is described in~\cite{DarcyCompile}. Throughout, we assume flows are either laminar or turbulent, which is consistent with most water distribution systems~\cite{ReynoldCite}.  
Each pipe $(i,j)\in \mathcal{E}$ has a  length $l_{ij}$ and a diameter $d_{ij}$, which are constant. Moreover, each pipe has a Darcy friction constant $\varepsilon_{ij}$, which captures how easily water flows through the pipe. The flow through the pipe $(i,j)\in \mathcal{E}$ is described by the flow quantity $q_{ij}$ and the flow velocity $v_{ij}$, where
$$ q_{ij}=\frac{v_{ij}d_{ij}^2 \pi }{4}. $$
Indeed, $q_{ij} = -q_{ji}$. Since a leaf vertex $i$ only has a single  vertex $j$ as its neighbor, we denote $q_{ij}=q_i$ for each leaf vertex.
The flows are controlled with the pressure head $H_{i}$, for each $i\in \mathcal{V}$. It can be shown that the flows through each junction $i\in \mathcal{V}$ are conserved 
\begin{proposition}\label{prop:ComeInComeOut}
    For an incompressible fluid, each 
    junction $i$ with water demand $D_i$, must fulfill
    \begin{equation}\label{eq:ComeInComeOut}
        \textstyle \sum_{j\in \mathcal{V}_i} {q_{ji}} = D_i.
    \end{equation}
\end{proposition}
%
%

The difference in pressure head between two 
junctions $i$ and $j$ is called head loss. Head loss is related to the volume flow of the liquid through the empirical Darcy-Weisbach equation,
\begin{equation}\label{eq:BasicHeadLoss}
    H_i-H_j=f(q_{ij},\varepsilon_{ij}, d_{ij})\frac{8 l_{ij} }{d_{ij}^5\pi^2g }{q_{ij}}\abs{{q_{ij}}}+m_{ij}{q_{ij}}\abs{{q_{ij}}},
\end{equation}
where $f(q_{ij},\varepsilon_{ij}, d_{ij})$ is the Darcy friction factor and $m_{ij}$ is the minor loss, a constant capturing the disturbance on the flow from valves and turns~\cite{EPANET}. 
By defining \begin{equation}\label{eq:Udef}
\begin{aligned}
        U_{ij}&=\frac{8 f(q_{ij},\varepsilon_{ij}, d_{ij}) }{d^5_{ij}\pi^2 g }{q_{ij}}\abs{{q_{ij}}}\\
        M_{ij}&=m_{ij}{q_{ij}}\abs{q_{ij}},  
\end{aligned}
\end{equation} equation~\eqref{eq:BasicHeadLoss} can be written as,
\begin{equation}\label{eq:UHeadloss}
    H_i-H_j=l_{ij} U_{ij}+M_{ij}
\end{equation}

Finally, a leakage is traditionally modeled as a demand {which is proportional to the pressure head~\cite[Equation~(1)]{leakJournal}.} For example, a leakage at vertex $i$ is given as,
\begin{equation}\label{eq:leak}
    {D_i}=C_i H_i^{\beta_i},
\end{equation}
where $C_i$ and $\beta_i$ are, respectively, the \emph{leak constant} and \emph{leak exponent}, at vertex $i$. The parameter $C$ is related to the size, while $\beta$ describes the severity of the leak. The parameter $\beta$ goes from small to large when the leak goes from hole to a break on the pipe.  Throughout the paper, we model a leak inside a pipe $(i,j)\in \mathcal{E}$ by introducing an auxiliary vertex between vertices $i$ and $j$ with the demand profile in~\eqref{eq:leak}.

This paper introduces two new quantities, apparent volume flow (denoted as $q^*$) and apparent pressure head (denoted as $H^*$). Both these notions rely on the assumption that there are no unidentified leaks in the distribution system. 
Apparent volume flow, $q^*_{ij}$ is the volume flow between vertices $i$ and $j$. 
Apparent pressure head, $H^{*}_{ji}$ is the pressure head at vertex $i$ calculated using the pressure head at vertex $j$.


\subsection{Problem Formulation: Leakage Localization}\label{section:Assumption}

 Our goal is to localize leakages in the water distribution system. In current water distribution systems, it is usually only possible to take measurements from the leaf vertices $\mathcal{L}$, i.e., the providers (\sebcancel{for instance, storage tanks, water treatment plants, water utility companies}), the end-users, and at certain junctions. Therefore, we focus on how we can localize leakages by using only measurements from leaf vertices. In particular, we assume that we have measurements of the pressure $H_i$ and flow $q_{i}$ for each leaf vertex $i\in \mathcal{L}$. 
 
 \seb{Throughout the paper, we assume that
the leaf vertices $\mathcal{L}$ are exactly the points in the water distribution network where water enters or exits the system.
 Given such an assumption,}
 it is easy to detect whether (or not) there is a leak in the system, by checking if the inflow matches the outflow. If the inflow matches the outflow, then clearly there is no leak in the system; otherwise, we say that there is at least one leak in the system. The case where there is no leak in the system is trivial, and, consequently, of no interest in the present paper. Therefore, throughout the rest of the paper we make the following assumption.
 \begin{assumption}\label{assump:1}
There is one leak in the water distribution system. Furthermore,  the leak does not appear on a pipe section with minor losses (i.e., for $m=0$).
\end{assumption}
We can now introduce an artificial junction $\lambda$ on the pipe $(a,b)\in \mathcal{E}$ where  the leakage is.
 The central goal of the leakage detection problem is to find the pipe $(a,b)$ with the leak and the distance to leak from vertex $b$
  in the said 
 pipe. 
 By the fluid dynamics model given in \ref{sec:WaterDistributionSystems}, this task reduces to finding 
 a solution for $x$ and $(a,b)$ to the following system of equations:
\begin{align}
    &H_{i}-H_{j}=l_{ij} U_{ij}+M_{ij}  &&\forall(i,j)\in \mathcal{E}\setminus \{(a,\lambda),(\lambda,b)\} \nonumber\\
    &\textstyle\sum_{j\in \mathcal{V}_i}q_{ij}=0 &&\forall  i \in \mathcal{V}\setminus (\mathcal{L}\cup \{\lambda\}) \nonumber\\
    &H_{b}-H_{\lambda}=x U_{b\lambda} \label{eq:SetEquations} \\
    &H_{\lambda}-H_{a}=(l_{ab}-x) U_{\lambda a} \nonumber \\
    &D_\lambda=q_{b\lambda}-q_{\lambda a}=CH_\lambda^{\beta}.  \nonumber
\end{align}  
 In the rest of this paper,
 we 
 prove that the solution to these equations  is unique, and we provide an algorithm to find the pipe $(a,b)$ and the location $x$.

\section{Leak localization -Single Pipe} \label{sec:single pipe}

The simplest non-trivial system is a homogeneous pipe with no minor losses. It can described as $\mathcal{T}=(\mathcal{V},\mathcal{E})$, for which $\mathcal{V}=\{0,\lambda,1\}$ and $\mathcal{E}=\{(0,\lambda),(\lambda,1)\}$, where $\lambda$ is the vertex with the leak. For this system equation~\eqref{eq:SetEquations} becomes,

\begin{equation}\label{eq:SetEquationsSimple}
        \begin{cases}
            H_{0}-H_\lambda=x U_{0\lambda }\\
            H_\lambda-H_{1}=(l_{01}-x) U_{\lambda 1}\\
            D_\lambda=q_{0}+q_{1}=CH_\lambda^\beta
        \end{cases}
\end{equation}

\begin{theorem}\label{th:sen1compile}

    Consider the system of equations~\eqref{eq:SetEquationsSimple}. If $H_0$, $H_1$, $q_0$, $q_1$ are known, then~\eqref{eq:SetEquationsSimple} 
    has a unique solution for $x$ which is given by,
    
     
     
     \begin{equation}\label{eq:sen1x}
         x =\frac{H_{0}-H_1-l_{01}U_{\lambda 1}}{U_{0\lambda }-U_{\lambda 1}}.
    \end{equation}
    Furthermore, if there is another measurement, 
    $H'_0$, $q'_0$, and $q'_1$, such that $H_0\neq H'_0$, and $q_0+q_1\neq q'_0+q'_1$, then $\beta$ and $C$ are uniquely determined as,
    \begin{equation}\label{eq:sen1bc}
       \begin{aligned}
         \beta =\frac{\log\left(\frac{D_\lambda}{D_\lambda'}\right)}{\log\left(\frac{H_0-x U_{0\lambda }}{H'_0-x U'_{0\lambda }}\right)},
         \,
         C =\frac{D_\lambda}{(H_0-x U_{0\lambda })^{\beta}}\hspace{1mm}.
        \end{aligned}
    \end{equation}

\end{theorem}

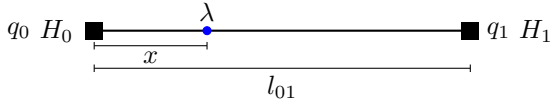
\begin{figure}[H]
    \centering
    \begin{center}

\begin{tikzpicture}[scale=0.5]
    \draw[thick] (-5,0) to (5,0);
    \filldraw[blue] (-2,0) circle (3pt);
    \node[rectangle,
    fill = black] (r) at (-5,0) {};
    \node[rectangle,
    fill = black] (r) at (5,0) {};
    \node[black] at (-6,0) {$H_0$};
    \node[black] at (-7,0) {$q_0$};
    \node[black] at (5.75,0) {$q_1$};
    \node[black] at (6.75,0) {$H_1$};
    \node[black] at (-2,0.5) {$\lambda$};

    \draw (-5,-1) to (5,-1);
    \draw (-5,-1.1) to (-5,-0.9);
    \draw (5,-1.1) to (5,-0.9);
    \node at (0,-1.5) {$l_{01}$};
    
    \draw (-5,-0.4) to (-2,-0.4);
    \draw (-5,-0.3) to (-5,-0.5);
    \draw (-2,-0.3) to (-2,-0.5);
    \node at (-3.5,-0.7) {$x$};

\end{tikzpicture}    
        \caption{Two sensors (squares), connected by a pipe (line) with a leak (blue circle) at junction $\lambda$.}
        \label{fig:pipe1}
    \end{center}
\end{figure}

Theorem~\ref{th:sen1compile} 
provides a closed-form expression  for locating the leak on a pipe, by taking recourse to pressure measurements and flow measurements; see Figure~\ref{fig:pipe1} for a depiction of the setting.
In a water distribution system, the amount of water that enters and leaves through the two leaf junctions should change with time, thereby allowing for the leak constant and leak exponent to be determined. The said parameters 
are used for estimating the severity of the leak. \sebcancel{In particular, if, as previously mentioned, $\beta$ is very high, then the leak can be considered as a break/major crack in the pipe; small vales of $\beta$ would imply existence of a minor hole.}
While Theorem~\ref{th:sen1compile} considers a single leak on a single pipe, it is natural to encounter scenarios where more than one leak is present in a pipe. For a system with $N$ leaks, assuming
perfect knowledge 
of the location of the leak, leak constant, and leak exponent for each of the other $N-1$ leaks, one could use Theorem~\ref{th:sen1compile} to localize the $N$th leak. 

In practice, sensors are 
oftentimes imperfect, and, consequently, provide faulty measurements. In the rest of this section, we focus on leak localization in the presence of  inaccurate and imprecise measurements. 
One approach towards modeling imprecision in sensor measurements is to assume 
that the measurements are $\hat{H}_i=H_i+\Delta_{H_i}$ and $\hat{q}_{ij}=q_{ij}+\Delta_{q_{ij}}$, where $\Delta_{*}$ is a constant offset of quantity $*$. Using this assumption the system of equations~\eqref{eq:SetEquationsSimple} are rewritten as  

\begin{equation}\label{eq:SetEquationsSimpleOffset}
      \begin{cases}
            \hat{H}_{0}-\hat{H}_\lambda=x \hat{U}_{0\lambda }\\
            \hat{H}_\lambda-\hat{H}_{1}=(l_{01}-x) \hat{U}_{\lambda 1}\\
            \hat{D}_\lambda=\hat{q}_{0}+\hat{q}_{ 1}=C\hat{H}_\lambda^\beta 
        \end{cases}
\end{equation}

\begin{proposition}\label{th:offset}
Consider the system of equations in \eqref{eq:SetEquationsSimpleOffset}. Let $\hat{H}_0$ (resp. $\hat{H}_1$) and $\hat{q}_0$ (resp. $\hat{q}_1$) denote the pressure measurement at vertex $0$ (resp. vertex $1$), and flow measurements at vertex $0$ (resp. vertex $1$), respectively. Let $\Delta=(\Delta_{H_0},\Delta_{H_1},\Delta_{q_{0\lambda}},\Delta_{q_{\lambda 1}})$ denote the offset in the measurements. Then, the estimate of the leak location is given as    $\hat{x}=x+\Delta_x$, where $\Delta_x$ is,
\begin{equation}\label{eq:DeltaX}
    \begin{aligned}
         \Delta_x&=\frac{\Delta_{ H_{0}}-\Delta_{H_{1}} - x \frac{\partial U_{0\lambda }}{\partial q_{0\lambda }} \Delta_{ q_{0\lambda }}+(x-l_{01})\frac{\partial U_{\lambda 1}}{\partial q_{\lambda 1}}\Delta_{q_{\lambda 1}}}{U_{0\lambda }-U_{\lambda 1}}\\
         &+\underbrace{\frac{\partial^2 x}{\partial H_0^2}\frac{\Delta_{H_0}^2}{2}+\frac{\partial^2 x}{\partial H_1^2}\frac{\Delta_{H_1}^2}{2}+...}_{\text{H.O.T.}}
         \end{aligned}
     \end{equation}
\end{proposition}
In words, Proposition~\ref{th:offset} says that for the case of a single leak in a system with only one pipe,  given pressure and flow measurements that are possibly offset, 
it is possible to obtain an estimate of the location of the leak. The error in the said estimate depends on the quantity on the right-hand side of~\eqref{eq:DeltaX}. If it is too large, then no conclusions can be drawn regarding the estimate $\hat{x}$; if it is quite small, then the estimate $\hat{x}$ is a good one.\\
Next, we consider the case where the pressure and flow measurements are possibly corrupted by noise. 
We have the following result.


\begin{theorem}\label{th:sen1Noise}
   Consider the system of equations \eqref{eq:SetEquationsSimpleOffset}. Let $\hat{H}_0$ (resp. $\hat{H}_1$) and $\hat{q}_0$ (resp. $\hat{q}_1$) denote the pressure measurement at vertex $0$ (resp. vertex $1$), and flow measurements at     vertex $0$ (resp. vertex $1$), respectively. Assume that the noise, in each of the measurements, is Gaussian and that it is independent.
   Let $\sigma^2=(\sigma_{H_0}^2,\sigma_{H_1}^2,\sigma_{q_{0 \lambda}}^2, \sigma_{q_{\lambda 1}}^2)$ denote the variance in the measurements.
   Then, the estimate of the leak location is given as $\hat{x}=x+\Delta_x$, where, \seb{assuming that $\norm{\sigma^2}$ is sufficiently small,} $\Delta_x\sim \mathcal N(0,\sigma_x^2)$,
   with 

    \begin{equation}\label{eq:sen1Var}
    \begin{aligned}
        \sigma_x^2&=\frac{\sigma_{H_0}^2+\sigma_{H_1}^2+x^2 \frac{\partial U_{0\lambda }}{\partial q_{0\lambda }}^2 \sigma^2_{q_{0\lambda }}+(x-l_{01})^2\frac{\partial U_{\lambda 1}}{\partial q_{\lambda 1}}^2\sigma_{q_{\lambda 1}}^2}{\left(U_{0\lambda }-U_{\lambda 1}\right)^2}\\
        &+\underbrace{\frac{\partial^2 x}{\partial H_0^2}^2\frac{\sigma_{H_0}^4}{4}+\frac{\partial^2 x}{\partial H_1^2}^2\frac{\sigma_{H_1}^4}{4}+...}_{\text{H.O.T.}}.
        \end{aligned}
    \end{equation}  
\end{theorem}






The theorem shows that if there is noise in the measurements with variance $\sigma^2$ then the estimation of the leakage will have variance $\sigma_x^2$ (up to H.O.T.). This means that if $\sigma^2$ is small then the mean square error of the estimation goes to zero as the number of measurements grows to infinity. For example, if we have  $N$ i.i.d. measurements, then we get a \seb{$95\% $} confidence bound of $[\hat{x}-1.96\frac{\sigma_x}{\sqrt{N}},\hat{x}+1.96\frac{\sigma_x}{\sqrt{N}}]$. 

\section{Leak localization - Distribution Tree}\label{sec:tree}
Theorem~\ref{th:sen1compile} concerns localization of a single leak in a system of single pipe. However, modern water networks are far more complex. In particular, one often encounters water distribution systems with more than one pipe. Finding the location of a leak in such systems is more challenging than the single pipe case discussed in Section~\ref{sec:single pipe}. As such, in this section, we detail when and how the location of a leak in a water distribution system (comprising of two or more pipes) can be obtained. \\

It turns out that Theorem~\ref{th:sen1compile} can be generalized to account for the presence of more than one pipe. In particular, if the water distribution network has a tree topology, then with sufficient knowledge of the system the leak can be located and the corresponding leak constant and leak exponent calculated via an algorithm. First, we need the following Proposition and Lemma.

\begin{proposition}\label{prop:DistTreeStep}
    Consider distribution tree $\mathcal{T}=(\mathcal{V},\mathcal{E})$ where all the minor losses are known. For any two neighboring vertices $s,t\in \mathcal{V}$, 
    let $\mathcal{L}_{st}=\{i\in\mathcal{L} |t \notin (k)_{k=i}^{s}\}$. Define the (sub)tree $\mathcal{T}_{st} (\subseteq \mathcal{T})=(\mathcal{V}_{st},\mathcal{E}_{st})$ 
    such that $\mathcal{L}_{st}\cup \{t\}\subseteq\mathcal{V}_{st}$. Suppose that there are no leaks in $\mathcal{T}_{st}$ and that for each $i\in\mathcal{L}_{st}$, $H_i$ and $q_i$ are known. For a leaf vertex $p\in \mathcal{L}_{st}$, $H_s$ and $q_{st}$ are thus determined by 
    \begin{equation}\label{eq:Ustep}
        \begin{aligned}
            q_{st}&=\textstyle\sum_{i\in \mathcal{L}_{st}}q_{i}\\
            H_{s}&=H_{p}-\textstyle\sum_{i\in (k)_{k=p}^{s-1}}\left(l_{i(i+1)} U_{i(i+1)}-M_{i(i+1)}\right)
        \end{aligned}
    \end{equation}
    
\end{proposition}
\begin{lemma}\label{lem:lowApp}
    Consider a tree $\mathcal{T}$ that contains a single leak. The apparent pressure at vertex $a\in \mathcal V$     calculated from a leaf in \seb{a} 
    subtree of vertex $w$ that contains the leak is lower than that for a leaf in any other subtree of vertex~$w$.
\end{lemma}

\noindent With Lemma~\ref{lem:lowApp} in place, we have the following theorem.
\begin{theorem}\label{th:TreeIsolate}
    Consider
    the system of equations~\eqref{eq:SetEquations} under Assumption~\ref{assump:1}. 
    Suppose that the underlying graph of the water distribution network is a tree.  Suppose that, for each leaf $i\in\mathcal{L}$, $H_i$ and $q_i$ are known. Then  $x$, $\beta$, and $C$ can be uniquely determined by Algorithm~\ref{alg:TreeLoc}.
\end{theorem}

Algorithm \ref{alg:TreeLoc} 
arbitrarily picks  a non-leaf vertex from the distribution tree $\mathcal T$. Since we do not allow for leaves to be picked, any vertex  picked must have at least two neighbors.
Subsequently,
 Algorithm~\ref{alg:TreeLoc} looks at all the subtrees (where a subtree is as defined in Section~\ref{sec:notations}) with the said 
 vertex, and compares the apparent pressure at the aforementioned vertex from at least one leaf vertex in each of the subtrees. 
We know from Lemma~\ref{lem:lowApp} that the  subtree with the lowest apparent pressure contains the leak; call this the leaking subtree. The leaking subtree defines a new distribution tree. 
On this new distribution tree, a non-leaf vertex is chosen again arbitrarily. The aforementioned process is repeated on this new distribution tree. Since a non-leaf vertex has at least two neighbors, the cardinality of the leaking subtree is always smaller than the cardinality of the original (sub)tree. This means that at each iteration the cardinality of the leaking subtree is decreasing.
These steps are repeated  until only the leaking pipe is left. 
Subsequently, by using the result in Theorem~\ref{th:sen1compile}
the leak's position, exponent, and constant can be uniquely determined. \\
Observe that no matter which vertex is chosen, none of the  vertices are picked more than once. Therefore, the total number of times the vertices in $\mathcal T$ are picked does not exceed $F$. 
After a vertex is picked  the apparent pressure is calculated for a leaf in each subtree (that contains the said vertex). We then look for the subtree that has the lowest apparent pressure. 
To do so, we can use a line search, which, since there are $F$ vertices in $\mathcal T$, has complexity $\mathcal{O}(F)$.
Thus the outer loop  has complexity $\mathcal{O}(F)$, and the inner loop has complexity $\mathcal{O}(F)$; hence the complexity of Algorithm~\ref{alg:TreeLoc} is $\mathcal{O}(F^2)$. 


Observe that  Theorem~\ref{th:TreeIsolate}  is restricted to networks with tree topology, and accounts for the presence of only one leak. Scenarios more general than this, such as the presence of multiple leaks, water distribution networks with arbitrary topology (not necessarily tree) are beyond the scope of the present paper and are the  foci of ongoing investigations. Notice also that the setting in Theorem~\ref{th:TreeIsolate} does not account for noisy pressure and/or flow measurements. 


\begin{algorithm}
\caption{Leak localization Algorithm in a distribution tree}
\label{alg:TreeLoc}
 \hspace*{\algorithmicindent} \textbf{Input:} A distribution tree $\mathcal{T}=(\mathcal{V},\mathcal{E})$,  with a leak. The distribution tree fulfil the system of equations~\eqref{eq:SetEquations}, and, for each leaf $i\in \mathcal{L}$, $H_i$ and $q_i$ are known.\\ 
 \hspace*{\algorithmicindent} \textbf{Output:} The leak's $x$,  $\beta$, and $C$. 
\begin{algorithmic}[1]
\State Define a temporary graph $\mathcal{T}'=(\mathcal{V}',\mathcal{E}')$ and set $\mathcal{T}'=\mathcal{T}$.
\While{$|\mathcal{V}'|>2$}
    \State Take a non-leaf vertex $a\in \mathcal{V}'$ and set $H^*=\infty$
    
    \For{For each vertex $b\in \mathcal{V}'_a$}
    \State Pick a leaf $i$ so that $b\in (k)_{k=i}^{a}$.
    Calculate~$H^*_{i a}$.
    \If{$H^*_{ia}<H^*$}
        \State Set $H^*=H^*_{ia}$ and $v_{\text{min}}=b$
    \EndIf
    \EndFor
    \State Take the subtree $\mathcal{T}''=(\mathcal{V}'',\mathcal{E}'')$ of $\mathcal{T}'$ where $\mathcal{V}''=\{j\in \mathcal{V}'|v_{\text{min}}\in\{k\}_{k=j}^{a}\}\cup\{a\}$.
    
    \State Use Proposition~\ref{prop:DistTreeStep} to calculate $H_{a}$ and $q_{av_{\text{min}}}$.
    \State Set $\mathcal{T}'=\mathcal{T}''$
\EndWhile

\State Use Theorem~\ref{th:sen1compile} to calculate leak's $x$, $\beta$ and $C$.

\end{algorithmic}
\end{algorithm}

\section{Simulations}\label{sec:simulations}
In this section, we illustrate our theoretical findings. To this end, we discuss several simulations, all of which are run using a water distribution system modeling software package, EPANET, developed by the US Environmental Protection Agency's Water Supply and Water Resources Division; see \cite{EPANET} for more details regarding EPANET.\\

\begin{figure}[h]
    \centering
    \includegraphics[width=1\linewidth]{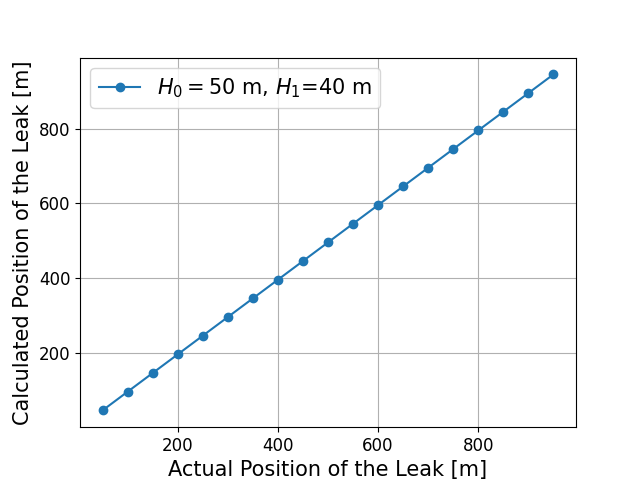}
    \caption{
    \seb{Position of the leak using Theorem \ref{th:sen1compile} in the single pipe system, against the position of the leak.}}
    \label{fig:OneLeakLocate}
\end{figure}

We consider a single pipe with a single leak setting. In particular, we choose a circular pipe circular with diameter $d=0.3\si{.m}$, length $l=1\si{.km}$, and Darcy friction constant $\varepsilon=0.15\si{.mm}$. Let us label the end-points of the pipes as vertices $0$ and $1$, respectively. We create leaks in increments of $100\si{.m}$ along the length of the pipe, and for each such leak (only one at a time) we measure the pressure head and volume flow at vertices $0$ and $1$, respectively. Based on these measurements, the position of the leak is computed using the result in Theorem~\ref{th:sen1compile}. 
We see that 
the actual position of the leak and the computed position of the leak coincide; see Figure~\ref{fig:OneLeakLocate}.

\seb{That the simulations very closely mirror the theoretical findings is not surprising, since EPANET also uses the Darcy-Weisbach equation to model the flow of water.}

\begin{figure}
    \centering
    \includegraphics[width=1\linewidth]{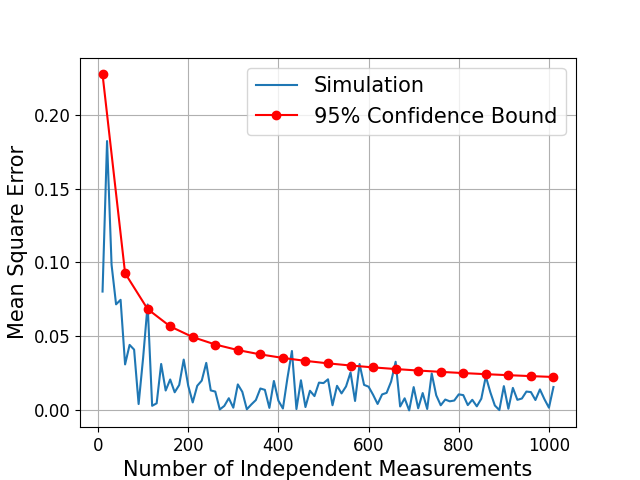}
    \caption{Mean square error against the number of independent measurements. 
    }
    \label{fig:OneLeakNoise}
\end{figure}

\section{Conclusion}\label{sec:conclusions}
In this paper, we considered water distribution networks and studied the problem of locating the position of a single leak across several scenarios. For the particular case wherein the water network just consists of a single leak and assuming the measurements are noiseless, we provided a closed-form expression for identifying the location of the leak, the leak constant, and the leak exponent. Subsequently, for the aforementioned setting,  we showed that, even when the pressure and flow measurements are noisy, it was possible to obtain an estimate of the location of the leak. Finally, considering a water distribution network comprising of more than one pipe, having a tree topology, and  the presence of a single leak,
we provided a general result to locate the position of the said leak. 

A key limitation with the present paper is that it accounts only for the presence of one leak. Hence, a possible direction of future research would be to account for the presence of multiple leaks in a water distribution system. Yet another line of investigation could revolve around accounting for water distribution networks with arbitrary topologies, and not necessarily tree  graphs.
\par \sebcancel{Yet another limitation of this work is that we assume perfect knowledge of the system parameters, in particular we assume that $m$ is set to zero, and, furthermore, have knowledge of the roughness coefficient of the pipe. Clearly, in practise, it is difficult to estimate the correct roughness coefficient of the pipes involved, and, consequently, the estimate of position of leak could be (negatively) impacted.  Further works could look at designing algorithms for leak detection that are also robust to variations in said parameters.}
\par \sebcancel{The present paper assumes that the graphs representing water networks have a tree topology. In practise, said networks could also be cyclic, i.e., have meshes in them. Accounting for such settings is yet another possible line of future investigation. }
\bibliography{ref}
 \section*{Appendix}




\subsection{Proof of Proposition~\ref{prop:ComeInComeOut}}\label{proof:ComeInComeOut} 
    Assume that the water is incompressible with density $\rho$ and that for each vertex $i$ the mass of water is constant, that is $\Delta \bar{m}_i=\bar{m}_a-\bar{m}_l=0$, where $\bar{m}_a$ and $\bar{m}_l$ are the mass of the fluid arriving and leaving at vertex $i$, respectively. Suppose that water enters vertex $i$ from the pipes, and that it leaves through the pipes or through a leak $D_i$.
    
    The mass of water arriving and leaving at vertex $i$ are given by $\bar{m}_a=\sum_{k\in\mathcal{V}^+_i} \rho {q_{ki}}$ and $\bar{m}_l=\sum_{k'\in\mathcal{V}^-_i} \rho {q_{ik'}}+\rho D_i$, where $\mathcal{V}^+_i=\{k\in\mathcal{V}_i |q_{ki}>0\}$ and $\mathcal{V}^-_i=\{{k'\in \mathcal{V}_i}|q_{ik'}\geq 0\}$. Note that $\{{k'}\}\cup\{k\}=\mathcal{V}_i$ and $\{{k'}\}\cap\{k\}=\varnothing$. This implies that,
    \begin{align}
        0&=\textstyle\sum_{j\in\mathcal{V}^+_i}\rho q_{ji}-\textstyle\sum_{j\in\mathcal{V}^-_i}\rho q_{ij}-\rho D_i \nonumber \\
        & =\rho(\textstyle\sum_{j\in\mathcal{V}^+_i} q_{ji}+\textstyle\sum_{j\in\mathcal{V}^-_i} q_{ji}- D_i)= \rho(\textstyle\sum_{j\in\mathcal{V}_i}-q_{ji}-D_i) \nonumber 
    \end{align}
    Therefore, we obtain the continuity equation \break $\sum_{j\in \mathcal{V}_i}{q_{ji}}=D_i$.

We need the following lemma for proving some of our results.

\begin{lemma}\label{lem:lessq}
    For a path $(k)_{k=0}^{n}$, where $\delta(k)=2$ for $k\notin\{0,n$\}, with  vertices $i,j$ such that $0<i<j<n$, \break 
      $ q_{j(j+1)} \leq q_{(i-1)i}$.
\end{lemma}

\textit{Proof:}
Let $(k)_{k=0}^{n}$ be a path where $\delta(k)=2$ for $k\notin\{0,n\}$, then for any vertex $0< i<n$, equation~\eqref{eq:leak} gives that, $q_{(i-1)i}-q_{i(i+1)}=D_{i}\geq 0\implies q_{i(i+1)}\leq q_{(i-1)i}$.
    For any two vertices, $i,j$; $0< i<j< n$, this implies that,
    \begin{equation}
        q_{j(j+1)}\leq q_{(j-1)j}\leq \ldots\leq q_{(i+1)(i+2)}\leq q_{(i-1)i}.
    \end{equation}

\subsection{Proof of Theorem~\ref{th:sen1compile}}\label{proof:sen1compile}
    Suppose that $\mathcal{T}=(\{0,\lambda,1\},\{(0,\lambda),(\lambda,1)\})$ which fulfills the set of equations in equation \eqref{eq:SetEquationsSimple}. Suppose that $H_0$, $H_1$, $q_{0}$ and $q_1$ are known. The set of equations in \eqref{eq:SetEquationsSimple} can be combined to obtain
    \begin{equation}
  \scriptsize      \begin{aligned}
            H_{0}-H_{1} &=x(U_{0\lambda }-U_{\lambda 1})+l_{01} U_{\lambda 1}\\
            \implies x &=\frac{H_{0}-H_{1}-l_{01} U_{\lambda 1}}{U_{0\lambda }-U_{\lambda 1}}.  
        \end{aligned}
    \end{equation}
    
    Furthermore, suppose that $H_0'$, $q_0'$ and $q_1'$ are known such that $H_0\neq H_0'$ and $q_0+q_1\neq q_0'+q_1'$, then $\beta$ is given as,
    \begin{equation}\label{eq:mellansSteg}
 \scriptsize       \begin{cases}
            D_\lambda=q_{0\lambda }+q_{\lambda 1}=C{H_\lambda}^{\beta} \\
            D_\lambda'=q_{0\lambda }'+q_{\lambda 1}'=C{H'_{\lambda}}^{\beta}
        \end{cases}\implies
            \frac{D_\lambda}{D_\lambda'}=\left(\frac{H_\lambda}{H'_\lambda}\right)^\beta,
    \end{equation}
    
    which implies that
    \begin{equation}
    \scriptsize
    \begin{aligned}
        \beta &=\frac{\log\left(\frac{D_\lambda}{D_\lambda'}\right)}{\log\left(\frac{H_\lambda}{H'_\lambda}\right)}=\frac{\log\left(\frac{D_\lambda}{D_\lambda'}\right)}{\log\left(\frac{H_0-x U_{0\lambda }}{H'_0-x U'_{0\lambda }}\right)}, \normalsize{\text{ and }} \nonumber \\
        C&=\frac{D_\lambda}{(H_0-x U_{0\lambda })^{\beta}} \nonumber.
    \end{aligned}
    \end{equation}

              

\subsection{Proof of Proposition \ref{th:offset}}\label{proof:offset}

    The set of equations \eqref{eq:SetEquationsSimpleOffset} can be rewritten as
    \begin{equation}\label{eq:xhat}
  \scriptsize     \hat{x}=\frac{\hat{H}_0-\hat{H_1}-l_{01}\hat{U}_{\lambda 1}}{\hat{U}_{0\lambda}-\hat{U}_{\lambda1}},
    \end{equation} Taylor expansion around $(H_0, H_1, q_0, q_1)$ yields: \vspace{-4mm}
    
    \begin{equation}\label{eq:xexpand}
      \scriptsize  \begin{aligned}
         \hat{x}&=x+\frac{\partial x}{\partial H_{0}}\Delta_{H_{0}}+\frac{\partial x}{\partial H_{1}}\Delta_{H_{1}}+ \frac{\partial x}{\partial q_{0}}\Delta_{q_{0}}+\frac{\partial x}{\partial q_{1}}\Delta_{q_{1}}\\
        &+\frac{\partial^2 x}{\partial H_0^2}\frac{\Delta_{H_0}^2}{2}+\frac{\partial^2 x}{\partial H_1^2}\frac{\Delta_{H_1}^2}{2}+\frac{\partial^2 x}{\partial q_1^2}\frac{\Delta_{q_1}^2}{2}+\frac{\partial^2 x}{\partial q_2^2}\frac{\Delta_{q_2}^2}{2}\\
        &+\frac{\partial^2 x}{\partial H_0\partial H_1}\Delta_{H_0}\Delta_{H_1}+\frac{\partial^2 x}{\partial H_0\partial q_0}\Delta_{H_0}\Delta_{q_0}\\
        &+\frac{\partial^2 x}{\partial H_0\partial q_1}\Delta_{H_0}\Delta_{q_1}+\frac{\partial^2 x}{\partial H_1\partial q_0}\Delta_{H_1}\Delta_{q_0}+\\
        &+\frac{\partial^2 x}{\partial H_0\partial q_1}\Delta_{H_0}\Delta_{q_1}+\frac{\partial^2 x}{\partial q_0\partial q_1}\Delta_{q_0}\Delta_{q_1}+...
        \end{aligned}
    \end{equation}

    where $\hat{x}=x+\Delta_x$ and

    \begin{equation}
  \scriptsize     \begin{aligned}
          \frac{\partial x}{\partial H_0}&=-\frac{\partial x}{\partial H_1}=\frac{1}{U_{0\lambda }-U_{\lambda 1}}\\ 
          \frac{\partial x}{\partial q_{0\lambda }}&=\frac{-x}{U_{0\lambda }-U_{\lambda 1}}\frac{\partial U_{0\lambda }}{\partial q_{0\lambda }}\\
         \frac{\partial x}{\partial q_{\lambda 1}}&=\frac{x-l_{01}}{U_{0\lambda }-U_{\lambda 1}}\frac{\partial U_{\lambda 1}}{\partial q_{1}}
        \end{aligned}
    \end{equation}

\subsection{Proof of Theorem~\ref{th:sen1Noise}}\label{proof:sen1Noise}
    The set of equations \eqref{eq:SetEquationsSimpleOffset} can be rewritten as equation~\eqref{eq:xhat}, whose Taylor expansion yields equation \eqref{eq:xexpand}. We assume that $\Delta_*\sim \mathcal{N}(0,\sigma_*^2)$ (i.e. Gaussian distributed) 
    Since for $\Delta_i\sim \mathcal{N}(0,\sigma_i^2)$ and $\Delta_j\sim \mathcal{N}(0,\sigma_j^2)$ and constants $a,b$ we have $a\Delta_i+b\Delta_j\sim \mathcal{N}(0,(a\sigma_i)^2+(b\sigma_j)^2)$ and $a\Delta_i\cdot\Delta_j\sim \mathcal{N}(0,a^2\sigma_i^2\sigma_j^2)$ we find that,
    \begin{equation} \scriptsize
    \begin{aligned}
        \sigma_x^2&=\frac{\sigma_{H_0}^2+\sigma_{H_1}^2+\hat{x}^2 \frac{\partial U_{0\lambda }}{\partial q_{0\lambda }}^2 \sigma^2_{q_{0\lambda }}+(\hat{x}-l_{01})^2\frac{\partial U_{\lambda 1}}{\partial q_{\lambda 1}}^2\sigma_{q_{\lambda 1}}^2}{\left(U_{0\lambda }-U_{\lambda 1}\right)^2}\\
        &+\frac{\partial^2 x}{\partial H_0^2}^2\frac{\sigma_{H_0}^4}{4}+\frac{\partial^2 x}{\partial H_1^2}^2\frac{\sigma_{H_1}^4}{4}+\frac{\partial^2 x}{\partial q_1^2}^2\frac{\sigma_{q_1}^4}{4}+\frac{\partial^2 x}{\partial q_2^2}^2\frac{\sigma_{q_2}^4}{4}\\
        &+\frac{\partial^2 x}{\partial H_0\partial H_1}^2\sigma_{H_0}^2\sigma_{H_1}^2+\frac{\partial^2 x}{\partial H_0\partial q_0}^2\sigma_{H_0}^2\sigma_{q_0}^2\\
        &+\frac{\partial^2 x}{\partial H_0\partial q_1}^2\sigma_{H_0}^2\sigma_{q_1}^2+\frac{\partial^2 x}{\partial H_1\partial q_0}^2\sigma_{H_1}^2\sigma_{q_0}^2+\\
        &+\frac{\partial^2 x}{\partial H_0\partial q_1}^2\sigma_{H_0}^2\sigma_{q_1}^2+\frac{\partial^2 x}{\partial q_0\partial q_1}^2\sigma_{q_0}^2\sigma_{q_1}^2+... 
    \end{aligned} \nonumber
    \end{equation}
\subsection{Proof of Proposition~\ref{prop:DistTreeStep}}\label{proof:prop:DistTreeStep}
 Suppose that $\mathcal{T}$ is a distribution tree where the minor losses are known. Define the subtree $\mathcal{T}_{st}=(\mathcal{V}_{st},\mathcal{E}_{st})$, as in Proposition~\ref{prop:DistTreeStep}. Suppose that for each vertex $i\in \mathcal{L}_{st}$, $q_{i}$ and $H_i$ are known and that there are no leaks in $\mathcal{T}_{st}$.

Take a leaf vertex $p$, and define the path $(k)_{k=p}^t$. Consider a set of adjacent vertices $a$ and $b$ on the path, then equation~\ref{eq:leak} gives $q_{ab}=\sum_{i\in \mathcal{V}_a\setminus\{b\}} q_{ia}$, by applying equation~\ref{eq:leak} on each of the terms we get that $q_{ab}=\sum_{i\in \mathcal{V}_a\setminus\{b\}}\sum_{j\in \mathcal{V}_i\setminus\{a\}} q_{ji}=...=\sum_{i\in \mathcal{L}_{ab}} q_{i}$. As such $U_{ab}$ can be calculated using equation~\ref{eq:Udef} for every set of adjacent vertices $a$ and $b$ on the path.

For each set of adjacent vertices $a$ and $b$ on the path equation~\ref{eq:BasicHeadLoss} gives $H_a-H_b=l_{ab}U_{ab}+M_{ab}$, summed over the who path this gives \break $H_{s}=H_p-\sum_{i\in (k)_{k=p}^{s-1}} \left(l_{i(i+1)}U_{i(i+1)}-M_{i(i+1)}\right)$.

\subsection{Proof of Lemma~\ref{lem:lowApp}} \label{lem:lowApp:proof}

    Suppose that $\mathcal{T}$ is a distribution tree with a single leak. Pick an arbitrary non-leaf vertex \seb{$w$}. Assume that there is a leak at location $\lambda$ on the pipe between vertices $c$ and $d$, where $c$ is closer to \seb{$w$} than $d$. The leaves can be divided into three groups:  $\mathcal{L}_{d}=\{i\in \mathcal{L}'|d\in(k)_{k=i}^m\}$ (i.e., the set of leaf vertices that have a path to vertex \seb{$w$} that involves vertex $d$.);  $\mathcal{L}_{o}=\{i\in \mathcal{L}'|w\in(k)_{k=i}^d\}$, 
        (i.e., the set of leaf vertices whose only paths towards each other contain vertex \seb{$w$}), and the rest are in the set $\mathcal{L}_{c}$. Note that $\mathcal{L}_{c}\cup\mathcal{L}_{d}$. 
    For leaves ${k_o}\in \mathcal{L}_o$, ${k_d}\in \mathcal{L}_d$, and ${k_c}\in \mathcal{L}_{c}$, based on Proposition~\ref{prop:DistTreeStep} we obtain \seb{$H^*_{k_ow}=H_w$}, $H^*_{k_dd}=H_{d}$, and $H^*_{k_cu}=H_{u}$ respectively, where $u$ is the vertex where paths \seb{$({k})_{k=k_c}^{w}$} ($k_c \in \mathcal L_c$) and \seb{$({k})_{k=c}^{w}$} first meet. Furthermore, using Lemma~\ref{lem:lessq}, \seb{$q_{\lambda c}<q_{d\lambda}$, and since higher flow of water leads to a larger drop in pressure, we have that $\partial U_{ij}/\partial q_{ij}>0$. Therefore, we obtain} 
    \begin{equation}
    \begin{aligned}
    H_{c}&=H_d-l_{d\lambda}U_{d\lambda}-l_{\lambda c}U_{\lambda c}\\ &>H_{k_dd}^*-l_{d\lambda}U_{d\lambda}-l_{\lambda c}U_{\lambda c}^*=H_{k_dc}^*.
    \end{aligned}
    \end{equation}
    For any two adjacent vertices \seb{$i,j\in ({k})_{k=c}^{w}$} such that $j$ is closer to \seb{$w$}, the set of equations \eqref{eq:SetEquations} yields \break  $q_{ij}= \sum_{n\in \mathcal{L}_{ij}\setminus{\mathcal{L}_{dc}}}q_n+q_{\lambda c}
    =\sum_{n\in \mathcal{L}_{ij}}q_n-D_\lambda<q_{ij}^*$, where $\mathcal{L}_{pq}$ is defined as in Proposition \ref{prop:DistTreeStep}. As such, from equation~\eqref{eq:UHeadloss}, $H_{j}=H_i-l_{ij}U_{ij}-M_{ij}>H_i-l_{ij}U_{ij}^*-M_{ij}^*=H_{ij}^*$, and since $H_{d}\geq H_{nd}^*$ for any $n\in \mathcal{L}_{de}$, where $e$ is the adjacent vertex to $d$ that is closer to \seb{$w$} (in the sense that there is a path from vertex $e$ to \seb{$w$} that contains fewer vertices than all paths from vertex $d$ to \seb{$w$}). 
        Hence, $H_{j}>H_{nj}^*$ for $n\in \mathcal{L}_{ij}$. Therefore, it follows that \seb{$H^*_{k_ow}>H_{k w}^*$} where $k\in \mathcal{L}_{c}\cup \mathcal{L}_{d} $ which is a subtree of \seb{$w$}.

\vspace{-2mm}
\subsection{Proof of Theorem~\ref{th:TreeIsolate}}\label{proof:TreeIsolate}
    Suppose that $\mathcal{T}$ is a distribution tree with a single leak. Assume that, for each leaf $i$ $\in \mathcal{T}$, $H_i$ and $q_i$ are known. Define a temporary tree $\mathcal{T}'=(\mathcal{V}',\mathcal{E}')$, whose set of leaves is denoted by $\mathcal{L}'$.\\ 
  Check whether $|\mathcal{V}'|=2$. If yes, then Theorem~$\ref{th:sen1compile}$ can be directly used to locate the leak. If not, then $|\mathcal{V}'| \geq 3$.
            Pick an arbitrary non-leaf vertex $a\in \mathcal{V}'$.
            For each vertex $i \in \mathcal L ^{\prime}$, calculate $H_{ia}^*$, 
and compare it to each other. Let $j \in \mathcal L ^{\prime}$ denote the leaf vertex with the lowest apparent pressure head. Therefore, from Lemma~\ref{lem:lowApp}, it follows that the subtree of $a$ with vertex $j$ in it must contain the leak. \\
 Observe that there must exist some vertex $k \in \mathcal L ^{\prime}$ that satisfies  $H_{ka}^* > H_{ja}^*$. Now construct a subtree $\mathcal{T}''=(\mathcal{V}'',\mathcal{E}'')$, where $\mathcal{L}''= \mathcal L_c \cup \mathcal L_d \cup \{a\} $, \seb{where $\mathcal L_c, \mathcal L_d$ are as defined in the proof of Lemma~\ref{lem:lowApp}}. Since $\lvert\mathcal L_o\rvert \geq 1$ (\seb{where $\mathcal L_o$ is as defined in the proof of Lemma~\ref{lem:lowApp}}), this implies that $\lvert\mathcal{V}''\rvert < \lvert\mathcal{V}'\rvert$.
 Set $\mathcal{V}'' = \mathcal{V}'$. Check whether $\lvert\mathcal{V}'\rvert =2$, and repeat the process. Since $|\mathcal{V}'|$ decreases after each iteration and $|\mathcal{V}'|<\infty$, there exists a finite number of iterations after which $|\mathcal{V}'|=2$, at which point using Theorem~\ref{th:sen1compile}, the unknown  $x$, $C$, and $\beta$ can be uniquely determined.

\end{document}